\documentclass[twocolumn,aps,prb,showpacs]{revtex4}
\usepackage{amssymb,amsfonts,amsmath}
\usepackage[T1]{fontenc}
\usepackage[latin1]{inputenc}
\usepackage{graphicx}

\begin{document}

\title{Phenomenological model for the remanent magnetization of dilute
quasi-one-dimensional antiferromagnets}

\author{A. P. Vieira}
\email{apvieira@if.usp.br}

\author{S. R. Salinas}
\email{ssalinas@if.usp.br}

\affiliation{Instituto de Física, Universidade de São Paulo\\
             Caixa Postal 66318, 05315-970 São Paulo, SP, Brazil}

\date{\today}

\begin{abstract}
We present a phenomenological model for the remanent magnetization at low
temperatures
in the quasi-one-dimensional dilute antiferromagnets
(CH\( _{3} \)NH\( _{3} \))Mn\( _{1-x} \)Cd\( _{x} \)Cl\( _{3} \)\( \cdot  \)2H\( _{2} \)O
and (CH\( _{3} \))\( _{2} \)NH\( _{2} \)Mn\( _{1-x} \)Cd\( _{x} \)Cl\( _{3} \)\( \cdot  \)2H\( _{2} \)O.
The model assumes the existence of uncompensated magnetic moments induced in
the odd-sized segments generated along the Mn\( ^{2+} \) chains upon dilution.
These moments are further assumed to correlate ferromagnetically after removal
of a cooling field. Using a (mean-field) linear-chain approximation and reasonable
set of model parameters, we are able to reproduce the approximate linear temperature
dependence observed for the remanent magnetization in the real compounds.
\end{abstract}

\maketitle

\section{Introduction}

At low temperatures, quasi-one-dimensional magnetic systems exhibit a wealth
of interesting behavior, such as dimensional crossover, \cite{smith68,dejonge75,wang97}
random quantum paramagnetism, \cite{nguyen96} order-by-disorder phenomena,
\cite{oseroff95,azuma97} and Griffiths phases, \cite{fisher95,young96} which
have motivated many experimental and theoretical investigations. In most of
these systems, three-dimensional (3D) order is eventually induced by interchain
interactions. Taking advantage of the many analytical results available for
one-dimensional model systems, this situation has been modeled in a variety
of ways. Most of the existent approaches are based on linear-chain approximations,
\cite{scalapino75,trudeau95,schulz96} which treat correlations along the chains
in an exact way, while introducing interchain couplings via effective fields.
These approximations have been quite successfully applied to pure systems, and
have given rise to generalized Ginzburg-Landau theories, \cite{scalapino75,mckenzie95}
which account for fluctuations. Also, they have been widely used to explain
disorder effects, \cite{imry75,hone75,schouten80,korenblit93,eggert02} which
are among the main topics of research on quasi-one-dimensional systems. 

In the present work we consider a class of quasi-one-dimensional compounds,
\cite{paduan98,becerra00} represented by (CH\( _{3} \)NH\( _{3} \))MnCl\( _{3} \)\( \cdot  \)2H\( _{2} \)O
(abbreviated MMC) and (CH\( _{3} \))\( _{2} \)NH\( _{2} \)MnCl\( _{3} \)\( \cdot  \)2H\( _{2} \)O
(abbreviated DMMC), consisting of localized spin systems in which the Mn\( ^{2+} \)
ions (spin \( S=\frac{5}{2} \)) lie along the crystalline \( b \) axis, forming
chains, and are antiferromagnetically coupled to each other by an intrachain
interaction \( J/k_{B} \) around \( 3 \) K. Magnetic susceptibility and specific
heat measurements \cite{simizu84} indicate the onset of 3D long-range order
at N\'{e}el temperatures \( T_{N}=4.12 \) K for MMC and \( T_{N}=6.36 \) K for
DMMC, with the magnetic moments aligning along the \( a \) axis. These temperatures
are compatible with an interchain interaction
\( \left| J_{\perp }\right| \sim \left| J\right| \times 10^{-2} \).
The character of this interaction is not reported in the literature. However,
owing to the behavior of the materials upon dilution with non-magnetic Cd\( ^{2+} \)
ions, it has been suggested (see below) that ferromagnetic interchain couplings
are present. At temperatures above \( T\sim 10 \) K, susceptibility results
are well described by a one-dimensional Heisenberg \( S=\frac{5}{2} \) model,
but at lower temperatures anisotropy effects (probably of dipolar origin) become
relevant. \cite{simizu84} Calculations based on a classical anisotropic Heisenberg
model, with parameters derived from experiments on DMMC, reinforce the importance
of anisotropy. \cite{schouten80} In particular, the behavior of the correlation
length along the chains is predicted to cross over from Heisenberg-like to Ising-like
as the temperature is lowered.

Substitution of small amounts of non-magnetic Cd\( ^{2+} \) for Mn\( ^{2+} \)
ions induces the appearance of a remanent magnetization \cite{paduan98,becerra00}
below \( T_{N} \), when samples are cooled in the presence of fields of a few
Oe directed along the easy axis. This remanent magnetization is observed to
vary linearly with temperature, except immediately below \( T_{N} \), where
demagnetization effects seem to be relevant. \cite{paduan98} Moreover, there
is an excess parallel susceptibility, which is in general associated with the
existence of uncompensated magnetic moments in odd-sized segments formed along
the chains upon dilution. \cite{dupas78} Apparently, the linear temperature
dependence of the remanent magnetization is of universal character, as inferred
from measurements \cite{becerra00} performed on DMMC doped with Cd\( ^{2+} \)
(non-magnetic) and Cu\( ^{2+} \) (\( S=\frac{1}{2} \)). Experiments \cite{carvalho01}
performed on similar compounds, CsMnCl\( _{3} \)\( \cdot  \)2H\( _{2} \)O
(CMC) and CsMnBr\( _{3} \)\( \cdot  \)2H\( _{2} \)O (CMB) doped with Cu\( ^{2+} \)
(for which the signs of the interchain interactions are well known),
revealed that a remanent magnetization appears in CMB, in which interchain couplings
are ferromagnetic along one of the transversal directions and antiferromagnetic
along the other; in contrast, no net magnetization is observable in CMC, where
all interactions are antiferromagnetic. These experimental results, combined
with the observation that some effective ferromagnetic coupling is expected in
order to sustain a net remanent magnetization, have led to the idea that
ferromagnetic interchain interactions should also be present in DMMC and MMC.
\cite{becerra00} However, in the lack of experimental data, up to now no
conclusive evidence on this point seems to be available.

In this paper we introduce and discuss a phenomenological model for the
low-temperature magnetic behavior of those compounds.
By virtue of the previously discussed anisotropy effects, we believe the
qualitative aspects to be captured by a \( S=\frac{5}{2} \) Ising model, which
in the pure limit (and in the simplest case) is described by the Hamiltonian
\begin{equation}
\label{hamr}
{\cal H}=J\sum _{\boldsymbol {r}}S_{\boldsymbol {r}}S_{\boldsymbol {r}+\boldsymbol {b}}+\sum _{\boldsymbol {r}}\sum _{\boldsymbol {\delta }}J^{\perp }_{\boldsymbol {\delta }}S_{\boldsymbol {r}}S_{\boldsymbol {r}+\boldsymbol {\delta }},
\end{equation}
where \( J>0 \), \( \boldsymbol {r} \) is a lattice vector, \( \boldsymbol {b} \)
is the primitive vector along the crystalline \( b \) direction, \( \boldsymbol {\delta } \)
is a vector connecting a site to its nearest neighbors in the \( ac \) plane,
\( J_{\boldsymbol {\delta }}^{\perp }=J_{\perp }>0 \) if \( \boldsymbol {\delta } \)
is parallel to the \( a \) axis, and \( J_{\boldsymbol {\delta }}^{\perp }=-J_{\perp } \)
if \( \boldsymbol {\delta } \) is parallel to the \( c \) axis. Our approach
is based on a linear-chain approximation, which treats the intrachain couplings
(\( J \)) exactly, while introducing the weak interchain interactions (\( J_{\perp }\ll J \))
via Curie-Weiss terms connecting all spins (in such a way that a staggered effective
field results, combining both ferro- and antiferromagnetic interchain interactions
in a cooperative manner). At very low temperatures the chains are antiferromagnetically
ordered, with a characteristic two-sublattice structure. Upon dilution, a very
long chain breaks into finite segments, and uncompensated magnetic moments appear
at the ends of odd-sized segments. On phenomenological grounds, we assume these
moments to correlate ferromagnetically, with their direction determined in the
experiments by the cooling field. For each segment of spins, the partition function
can be exactly calculated; the total free energy of the chain is obtained by
summing the free energies of segments of all sizes, with proper weighting factors.
This procedure is detailed in Sec. \ref{sec2a}. Then, in Sec. \ref{sec2b},
we include the Curie-Weiss terms and discuss the results of the approximation.
We show that this approach reproduces the linear temperature dependence of the
remanent magnetization and the existence of an excess susceptibility. The final
section is devoted to a discussion and conclusions.

\section{Phenomenological model}

\subsection{\label{sec2a}Nearest-neighbor interactions}

We initially consider an open segment of \( n \) Ising spins with antiferromagnetic
couplings and alternating fields, described by the Hamiltonian

\begin{equation}
\label{hn}
{\cal H}_{n}=J\sum ^{n-1}_{j=1}S_{j}S_{j+1}-\sum ^{n}_{j=1}h_{j}S_{j}-D\sum ^{n}_{j=1}S^{2}_{j},
\end{equation}
where \( J>0 \) and \( h_{j}=h_{1} \) \( (h_{2}) \) for odd (even) \( j \);
a crystal field \( D \) is also introduced.  The spin variables \( S_{j} \)
take the values \( \pm \frac{1}{2} \), \( \pm \frac{3}{2} \) and \( \pm \frac{5}{2} \).
The alternating fields are introduced to give room to a staggered effective
field needed to describe long-range antiferromagnetic order in the presence
of interchain interactions. According to the phenomenological hypothesis that
there are uncompensated magnetic moments pointing in a preferred direction,
determined by the cooling field, we assume that the spins at the ends of odd-sized
segments are always under the action of a field \( h_{1} \). When the field
is removed, the moments would remain uncompensated due to pinning by the non-magnetic
impurities. For even-sized segments, the particular choice of a field \( h_{1} \)
at \( j=1 \) is of no consequence, since in this case the partition functions
are symmetric under inversion.

As we are considering finite values of \( n \), we must treat separately the
cases of odd- and even-sized segments. Using the transfer-matrix technique,
we can write the partition functions as
\begin{equation}
{\cal Z}_{n-1}^{\mathrm{odd}}=\left\langle \boldsymbol {v}_{1}\left| {\mathbf{T}}^{\frac{n-2}{2}}\right| \boldsymbol {v}_{1}\right\rangle
\end{equation}
and
\begin{equation}
{\cal Z}^{\mathrm{even}}_{n}=\left\langle \boldsymbol {v}_{1}\left| {\mathbf{T}}^{\frac{n-2}{2}}\mathbf{T}_{1}\right| \boldsymbol {v}_{2}\right\rangle =\left\langle \boldsymbol {v}_{2}\left| \mathbf{T}_{2}{\mathbf{T}}^{\frac{n-2}{2}}\right| \boldsymbol {v}_{1}\right\rangle ,
\end{equation}
 where \( n \) is an even number, \( \mathbf{T}=\mathbf{T}_{1}\mathbf{T}_{2} \),
the elements of the \( 6\times 6 \) matrices \( \mathbf{T}_{1} \) and \( \mathbf{T}_{2} \)
are given by
\begin{align}
T_{1}\left( S_{i},S_{j}\right)  & = e^{ -\beta JS_{i}S_{j}+
\frac{1}{2}\beta \left(h_{1}S_{i}+h_{2}S_{j}\right)+
\frac{1}{2}\beta D\left( S^{2}_{i}+S^{2}_{j}\right)} ,
\\
T_{2}\left( S_{i},S_{j}\right)  & = T_{1}\left( S_{j},S_{i}\right) ,
\end{align}
and the components of the vectors \( \boldsymbol {v}_{1} \) and \( \boldsymbol {v}_{2} \)
are
\begin{eqnarray}
v_{1}(S_{j}) & = & e^{\frac{1}{2}\beta \left( h_{1}S_{j}+DS^{2}_{j}\right) },\\
v_{2}(S_{j}) & = & e^{\frac{1}{2}\beta \left( h_{2}S_{j}+DS^{2}_{j}\right) }.
\end{eqnarray}
The free energies associated with odd- and even-sized segments are
\begin{equation}
F_{n-1}^{{\mathrm{odd}}}=-k_{B}T\ln {\cal {Z}}^{\mathrm{odd}}_{n-1}
\end{equation}
and
\begin{equation}
F_{n}^{{\mathrm{even}}}=-k_{B}T\ln {\cal {Z}}^{\mathrm{even}}_{n}.
\end{equation}

We now consider a very long chain, and assume that each of the \( N \) sites
is occupied by a spin with probability \( p \). For \( 0<p<1 \), the chain
is composed of finite segments of spins separated by empty sites. In the \( N\rightarrow \infty  \)
limit, the number of segments of size \( n \) is \( NP(n)=N(1-p)^{2}p^{n} \).
Assuming that each segment is described by the Hamiltonian in Eq. (\ref{hn}),
the total free energy per spin is given by the infinite series
\begin{equation}
f_{\mathrm{nn}}(h_{1},h_{2},T)=\frac{1}{p}\sum _{n\; \mathrm{even}}\left[ P(n-1)F^{\mathrm{odd}}_{n-1}+P(n)F^{\mathrm{even}}_{n}\right] .
\end{equation}
For \( p<1 \), as \( nP(n) \) becomes negligible for sufficiently large \( n \),
this infinite series can be truncated and readily evaluated numerically.

Let us denote by type 1 (type 2) those spins under the action of a field \( h_{1} \)
(\( h_{2} \)). The numbers \( N_{1} \) and \( N_{2} \) of spins of either
type in a chain can be determined by noting that in a segment of size \( n \)
there are \( n/2 \) type-1 spins if \( n \) is even and \( (n+1)/2 \) type-1
spins if \( n \) is odd. Thus, the fractions of type-1 and type-2 spins are
\begin{equation}
\frac{N_{1}}{N}=\sum _{n\; \mathrm{odd}}P(n)\frac{n+1}{2}+\sum _{n\; \mathrm{even}}P(n)\frac{n}{2}=\frac{p}{1+p},
\end{equation}
and 
\begin{equation}
\frac{N_{2}}{N}=\sum _{n\; \mathrm{odd}}P(n)\frac{n-1}{2}+\sum _{n\; \mathrm{even}}P(n)\frac{n}{2}=\frac{p^{2}}{1+p},
\end{equation}
 respectively. For \( p<1 \), the difference between the two fractions will
obviously generate a non-zero magnetization at zero temperature.

\subsection{\label{sec2b}Linear-chain approximation}

In order to mimic the weak interchain coupling in the real compounds, we now
assume that, in addition to the nearest-neighbor couplings inside each segment,
there are also ferromagnetic Curie-Weiss (CW) interactions connecting all spins
in the chain. We further assume that the CW interactions between two type-1
or two type-2 spins have strength \( J_{\mathrm{cw}}/N \), but that the CW
interactions between a type-1 and a type-2 spin are weaker by a factor \( \gamma  \).
This \( \gamma  \) factor is introduced to allow for off-plane interchain couplings;
in the pure limit (\( p=1 \)) the chains are expected to exhibit antiferromagnetic
order, so that \( \gamma  \) must be smaller than unity. Upon dilution, we
expect the antiferromagnetic arrangement to survive inside each segment, and
in principle this could lead to a variation of \( \gamma  \) with the concentration
\( p \), since the magnetic arrangement in the planes perpendicular to the
chains could be disturbed. In any case, our results suggest that \( \gamma  \)
is very small, if not zero, in the compounds under consideration. 

The contribution of the CW interactions to the total energy per spin is
\begin{equation}
\varepsilon _{\mathrm{cw}}=-pJ_{\mathrm{cw}}(m^{2}_{1}+2\gamma m_{1}m_{2}+m^{2}_{2}),
\end{equation}
 where \( m_{1} \) (\( m_{2} \)) is the magnetization per magnetic ion due
to spins of type 1 (type 2). Since \( \varepsilon _{\mathrm{cw}} \) depends
only on the averages \( m_{1} \) and \( m_{2} \), and not on the detailed
configuration of the spins, it is convenient to perform a change of variables.
So, we now introduce the Helmholtz potential per spin \( a_{\mathrm{nn}}(m_{1},m_{2},T) \),
related to the nearest-neighbor interactions, and defined by the Legendre transform
\begin{equation}
a_{\mathrm{nn}}(m_{1},m_{2},T)=f_{\mathrm{nn}}(\tilde{h}_{1},\tilde{h}_{2},T)+m_{1}\tilde{h}_{1}+m_{2}\tilde{h}_{2},
\end{equation}
where \( \tilde{h}_{1} \) and \( \tilde{h}_{2} \) are effective fields and
\begin{equation}
m_{1}=-\left( \frac{\partial f_{\mathrm{nn}}}{\partial \tilde{h}_{1}}\right) _{\tilde{h}_{2},T}\qquad {\mathrm{and}}\qquad m_{2}=-\left( \frac{\partial f_{\mathrm{nn}}}{\partial \tilde{h}_{2}}\right) _{\tilde{h}_{1},T}.
\end{equation}
For given values of \( m_{1} \) and \( m_{2} \) we can write the total Helmholtz
potential,
\begin{equation}
a(m_{1},m_{2},T)=a_{\mathrm{nn}}(m_{1},m_{2},T)+\varepsilon _{\mathrm{cw}},
\end{equation}
from which we obtain the relation between the external magnetic fields \( h_{1} \),
\( h_{2} \) and the effective fields,
\begin{equation}
h_{1}=\left( \frac{\partial a}{\partial m_{1}}\right) _{m_{2},T}=\tilde{h}_{1}-2pJ_{\mathrm{cw}}(m_{1}+\gamma m_{2}),
\end{equation}
and similarly
\begin{equation}
h_{2}=\left( \frac{\partial a}{\partial m_{2}}\right) _{m_{1},T}=\tilde{h}_{2}-2pJ_{\mathrm{cw}}(\gamma m_{1}+m_{2}).
\end{equation}
Comparing these last results (for \( \gamma =0 \)) with the local field at
a site \( \boldsymbol {r} \) due to its \( q_{\perp } \) nearest-neighbors
in adjacent chains, as given by the Hamiltonian in Eq. (\ref{hamr}), we conclude
that \( J_{\mathrm{cw}} \) can be estimated as
\begin{equation}
\label{jcwjp}
J_{\mathrm{cw}}\simeq \frac{1}{2}pq_{\perp }J_{\perp }.
\end{equation}

The thermodynamically stable magnetizations are those which minimize the free-energy
functional
\begin{eqnarray}
\Phi \left( h_{1},h_{2},T;m_{1},m_{2}\right)  & = & a(m_{1},m_{2},T)-m_{1}h_{1}-m_{2}h_{2}\nonumber \\
 & = & f_{\mathrm{nn}}\left( \tilde{h}_{1},\tilde{h}_{2},T\right) -\varepsilon _{\mathrm{cw}}.
\end{eqnarray}
For low temperatures and small ratios \( J_{\mathrm{cw}}/J \), setting \( h_{1}=h_{2}=0, \)
the stable values of \( m_{1} \) and \( m_{2} \) have opposite signs. In the
presence of dilution (\( p<1 \)), since we have \( \left| m_{1}\right| \neq \left| m_{2}\right|  \),
the model predicts a remanent magnetization per lattice site, \( m_{r} \),
given by
\begin{equation}
m_{r}=p(m_{1}+m_{2}).
\end{equation}
In the \( T\rightarrow 0 \) limit, \( m_{r} \) reaches a saturation value,
\begin{equation}
\label{mrsat}
\lim _{T\rightarrow 0}m_{r}=\frac{N_{1}-N_{2}}{N}S=\frac{p(1-p)}{(1+p)}S,
\end{equation}
where $S=\frac{5}{2}$.
The zero-field differential susceptibility \( \chi _{0} \) can be calculated
by setting \( h_{1}=h_{2}=h \) and taking the \( h\rightarrow 0 \) limit,
\begin{equation}
\chi _{0}=\lim _{h\rightarrow 0}\frac{\partial m_{r}}{\partial h}.
\end{equation}
The N\'{e}el temperature is obtained from the solution of the equation
\begin{equation}
\left\{ \frac{\partial ^{2}\Phi }{m^{2}_{1}}\frac{\partial ^{2}\Phi }{m^{2}_{2}}-\left( \frac{\partial ^{2}\Phi }{\partial m_{1}\partial m_{2}}\right) ^{2}\right\} _{m_{1}=m_{2}=0}=0,
\end{equation}
in the absence of an external field.

\begin{figure}
\includegraphics[angle=-90,width=7.8cm]{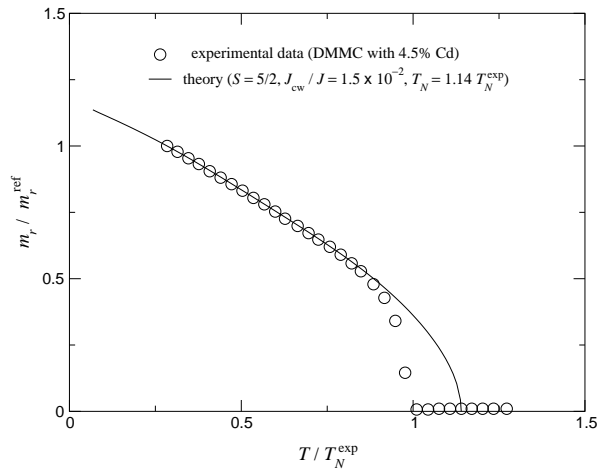}
\caption{
Experimental data (circles) and theoretical calculation (solid curve) for the
temperature dependence of the remanent magnetization in DMMC with \( 4.5\% \)
Cd. The magnetization is normalized to its value at the lowest temperature for
which experimental data are available.
\label{fig1}}
\end{figure}

In Fig. \ref{fig1} we show the experimental data \cite{becerra00} for the
temperature dependence of the remanent magnetization in DMMC doped with 4.5\%
Cd (this concentration was estimated from high-temperature fits to a Curie-Weiss
law). We also show results of our calculations for the remanent magnetization
with 4.5\% dilution, \( J_{\mathrm{cw}}/J=1.5\times 10^{-2} \), \( \gamma =0 \),
and \( D=0 \). We obtained the best fit for the linear portion of the curve
by setting the theoretical N\'{e}el temperature to 1.14 times the experimental value
(which amounts to fitting \( J \)). This is a reasonable procedure, since our
calculations are of a mean-field nature, and thus we do not expect to obtain
good quantitative results for the N\'{e}el temperature. Of course, the qualitative
features of the calculations are not sensitive to small variations in the parameters;
however, no strictly universal behavior (in the sense of data collapse) could
be identified. We point out that setting the value of the crystal field to high
positive values turns the system into a \( S=\frac{1}{2} \) Ising model, and
in this case the linear temperature dependence of \( m_{r} \) could not be
so well reproduced. Note that, in view of Eq. (\ref{jcwjp}), the value of \( J_{\mathrm{cw}}/J \)
used in the fit is fully compatible with the estimated experimental value of
\( J_{\perp }/J \) mentioned in the Introduction. The calculated ratio of the
N\'{e}el temperatures of the diluted and pure systems is \( 0.86 \), compared to
the experimental estimate \cite{becerra00} of \( 0.99 \) for the real material.
From Eq. (\ref{mrsat}), the saturation value of \( m_{r} \) for \( 1\% \)
dilution is \( 0.497\% \) of the sublattice magnetization in the pure system,
in excellent agreement with the experimental estimate \cite{paduan98} of \( 0.5\% \)
for MMC with \( 1\% \) Cd. 

\begin{figure}
\includegraphics[angle=-90,width=7.8cm]{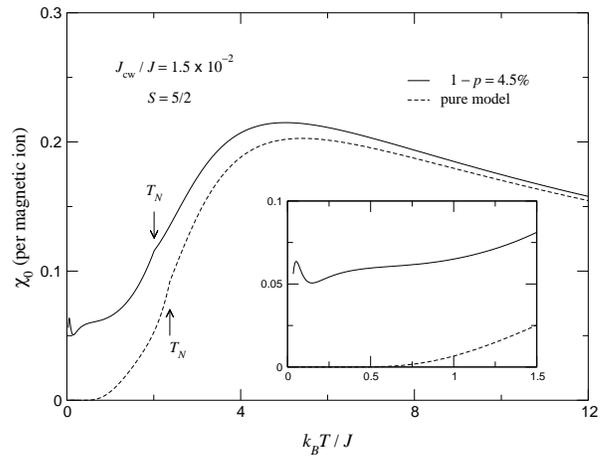}
\caption{
Calculated zero-field susceptibility per magnetic ion in the pure limit (dashed
curve) and for \( 4.5\% \) dilution (solid curve), using the same parameters
as in Fig. \ref{fig1}. The arrows indicate the corresponding N\'{e}el temperature,
which is lower in the dilute case. The inset shows low-temperature behavior.
\label{fig2}}
\end{figure}

In Fig. \ref{fig2} we use the previous set of parameters to plot the calculated
zero-field susceptibility \( \chi _{0} \) both in the pure limit and for \( 4.5\% \)
dilution. The broad maxima in the curves reflect short-range intrachain antiferromagnetic
correlations, while the cusps (indicated in the figure by the arrows) correspond
to the N\'{e}el temperatures (\( T_{N} \)). As it is clear in the inset, for the
dilute case we observe other features at lower temperatures. The small maximum
close to \( T=0 \) is due to isolated spins, whose sole energy scale is given
by the very weak interchain couplings, while the neighboring shoulder is due
to small odd-sized segments, whose end spins are uncompensated (even-sized segments
give negligible contributions to \( \chi _{0} \) at such low temperatures).
This is in sharp contrast to the pure limit, in which the susceptibility vanishes
exponentially for \( T<T_{N} \).

It should be mentioned that the present approach is a generalization of that
used by Slotte \cite{slotte85} to investigate the dilute \( S=\frac{1}{2} \)
Ising chain with competing short- and long-range interactions. However, owing
to the presence of competition, his approach does not contemplate the possibility
of long-range antiferromagnetic order at finite temperatures, even in the pure
limit.

\section{Conclusions}

We introduced a phenomenological model for the remanent magnetization (\( m_{r} \))
in a class of dilute quasi-one-dimensional antiferromagnets, composed of weakly
interacting spin chains. The model assumes the existence of uncompensated spins
at the ends of odd-sized segments formed along the chains upon dilution. These
spins are supposed to remain ferromagnetically correlated after a cooling field
is removed. By using a linear-chain approximation, in which interchain interactions
are treated at a mean-field level, we were able to reproduce the linear temperature
dependence of \( m_{r} \) for a set of parameters compatible with experimental
estimates. 

Our linear-chain approximation is based on the assumption that, even upon dilution,
each segment feels an staggered effective field. Of course, this assumption
is subject to some restrictions. Depending on the impurity concentration \( 1-p \),
the existence of uncompensated moments pointing in a preferred direction could
lead to a complete destabilization of the magnetic ordering perpendicular to
the chains (this can be seen by considering the effect, in a particular chain,
of two neighboring non-magnetic ions, which may invert the roles of the alternating
sublattices). In this case, spins along the chains would feel the same interchain
effective field, irrespective of their position. Actually, this would lead to
a ferromagnetic transition (with a diverging susceptibility), and the long-range
antiferromagnetic ordering would not be recovered even as \( p\rightarrow 1 \).
We have performed calculations near this limit, and have checked that the critical
temperature depends linearly on \( 1-p \), being thus too small compared with
experimental findings. Moreover, it is not possible to reproduce the linear
temperature dependence of the remanent magnetization. We conclude that, at least
in the low impurity concentrations used here, for which the occurrence of two
neighboring non-magnetic atoms in the same chain is a rare event, our approximation
is reasonable.

There remains the topic of identifying the precise mechanism responsible for
the persistence of ferromagnetic correlations between the uncompensated spins.
Monte Carlo simulations based on the Hamiltonian in Eq. (\ref{hn}) could be
used to verify whether it is sufficient or necessary to have both ferro- and
antiferromagnetic interchain interactions present in order to give rise to a
remanent magnetization in quasi-one-dimensional systems.

\begin{acknowledgments}
We thank C. C. Becerra, A. Paduan-Filho and T. A. S. Haddad for fruitful
discussions. This work was partially financed by the Brazilian agency CNPq and
Fapesp.
\end{acknowledgments}


\end{document}